\documentclass{article}%
\usepackage{amsfonts}
\usepackage{amsmath}
\usepackage{amssymb}
\usepackage{graphicx}%
\providecommand{\U}[1]{\protect\rule{.1in}{.1in}}
\providecommand{\U}[1]{\protect\rule{.1in}{.1in}}
\newtheorem{theorem}{Theorem}

\newtheorem{corollary}[theorem]{Corollary}

\newtheorem{definition}[theorem]{Definition}

\newtheorem{proposition}[theorem]{Proposition}
\newtheorem{remark}[theorem]{Remark}

\newenvironment{proof}[1][Proof]{\noindent\textbf{#1.} }{\ \rule{0.5em}{0.5em}}
\begin{document}

\title{St\"{a}ckel systems generating coupled KdV hierarchies and their finite-gap
and rational solutions}
\author{Maciej B\l aszak\\Department of Physics, A. Mickiewicz University\\Umultowska 85, 61-614 Pozna\'{n}, Poland\\blaszakm@amu.edu.pl
\and Krzysztof Marciniak\\Department of Science and Technology \\Campus Norrk\"{o}ping, Link\"{o}ping University\\601-74 Norrk\"{o}ping, Sweden\\krzma@itn.liu.se}
\maketitle

\begin{abstract}
We show how to generate coupled KdV hierarchies from St\"{a}ckel separable
systems of Benenti type. We further show that the solutions of these
St\"{a}ckel systems generate large class of finite-gap and rational solutions
for cKdV hierarchies. Most of these solutions are new.

\end{abstract}

\section{Introduction}

In \cite{bensol} we presented a systematic method of passing from St\"{a}ckel
separable systems to infinite hierarchies of commuting nonlinear evolutionary
PDE's. We presented the idea in a concrete case of St\"{a}ckel systems of
Benenti type. In this paper we recognize the obtained hierarchies as the well
known coupled Korteweg - de Vries (cKdV) hierarchies of Antonowicz and Fordy
\cite{AF} written in a different representation. We also clarify and
significantly simplify the approach developed in \cite{bensol}. The main idea
of the paper is however to present a new method of generating solutions of
soliton hierarchies from solutions of the related St\"{a}ckel systems.

From the very beginning of development of theory of integrable systems in the
late 60's major efforts have been put into constructing various ways of
finding their solutions. Among many others, a possible way of finding
solutions of integrable systems is through various kinds of symmetry
reduction, where one starts from an infinite-dimensional integrable system and
obtains after such reduction an integrable ODE. If one then succeeds in
solving this ODE (for example by finding separation coordinates in case of
Hamiltonian systems) one can then reconstruct the corresponding particular
solutions of the integrable PDE. This method originated in \cite{bogo} and has
been developed in \cite{AF}, \cite{antforstef}, \cite{stefmar}, \cite{skm},
\cite{cKdV} (see also \cite{mb1}) and later in a large number of papers. In
this paper we present an opposite approach in the sense that we start with
large classes of St\"{a}ckel systems written in separation coordinates so that
their solutions are explicitly known and in few steps we construct from these
St\"{a}ckel systems an infinite hierarchy of commuting evolutionary PDE's
while the solutions of the considered St\"{a}ckel systems become particular
multi-time solutions of the systems of the obtained hierarchy. In this way we
produce both well-known and new finite-gap type solutions of the KdV hierarchy
as well as new rational and finite-gap solutions of cKdV hierarchies. In many
cases this method also leads to implicit solutions of cKdV hierarchies.

The paper is organized as follows. In Section 2 we briefly describe the
starting point of our considerations, that is St\"{a}ckel separable systems of
Benenti type, including their general solution. In section 3 we relate with
our St\"{a}ckel systems a class of weakly-nonlinear semi-Hamiltonian systems
(i.e. a class of hydrodynamic-type systems) that are reductions of the so
called universal hierarchy \cite{al1}. These systems are in our formulation
defined by Killing tensors of St\"{a}ckel metrics. In section 4 we explicitly
construct hierarchies of commuting evolutionary PDE's and present a
transformation that maps these hierarchies onto the well known cKdV
hierarchies of Antonowicz and Fordy \cite{AF}. This section contains the main
result of this paper, i.e. Theorem \ref{glowny}, that produces large families
of solutions to our coupled hierarchies. Finally, section 5 is devoted to
studying specific classes of solutions that we call zero-energy solutions that
contain both rational and implicit solutions of our hierarchies.

\section{St\"{a}ckel systems of Benenti type}

Let us consider a set of canonical (Darboux) coordinates $(\lambda
,\mu)=(\lambda_{1}\ldots,,\lambda_{n},\mu_{1},\ldots,\mu_{1})$ on a
$2n$-dimensional Poisson manifold $M$. A set of $n$ relations of the form%

\begin{equation}
\varphi_{i}(\lambda_{i},\mu_{i},a_{1},\ldots,a_{n})=0\text{, \ }%
i=1,\ldots,n\text{, \ \ \ }a_{i}\in\mathbf{R}\label{SRalg}%
\end{equation}
(each involving only one pair $\lambda_{i},\mu_{i}$ of canonical coordinates)
are called separation relations \cite{Sklyanin} provided that $\det\left(
\frac{\partial\varphi_{i}}{\partial a_{j}}\right)  \neq0$. Resolving (locally)
equations (\ref{SRalg}) with respect to $a_{i}$ we obtain
\begin{equation}
a_{i}=H_{i}(\lambda,\mu),\text{ \ \ }i=1,\ldots,n.\label{hami}%
\end{equation}
with some new functions (Hamiltonians) $H_{i}(\lambda,\mu)$ that in turn
define $n$ canonical Hamiltonian systems on $M$:%
\begin{equation}
\lambda_{t_{i}}=\frac{\partial H_{i}}{\partial\mu},\ \ \mu_{t_{i}}%
=-\frac{\partial H_{i}}{\partial\lambda},\ \ \ \ i=1,...,n\label{2.2}%
\end{equation}
(here and in what follows the subscript denotes derivative with respect to the
subscript variable). From this setting it follows immediately that the
Hamiltonians $H_{i}$ Poisson commute. The corresponding Hamilton-Jacobi
equations for all Hamiltonians $H_{i}$ are separable in the $(\lambda,\mu
)$-variables since they are algebraically equivalent to the separation
relations (\ref{SRalg}).

In this article we consider a special but important class of separation
relations:%
\begin{equation}%
{\displaystyle\sum\limits_{j=1}^{n}}
a_{j}\lambda_{i}^{n-j}=Af(\lambda_{i})\mu_{i}^{2}+B\gamma(\lambda_{i})\text{,
\ \ \ }i=1,\ldots,n,\label{BenSR}%
\end{equation}
where $A$ and $B$ are two constants. Notice that since the functions $\gamma$
and $f$ do not depend on $i$ the relations (\ref{BenSR}) can in fact be
considered as $n$ copies of a curve - the so called \emph{separation curve} in
$\lambda$-$\mu$ plane. The Hamiltonian systems obtained from this class of
separation relations have been widely studied and are also known as Benenti
systems. Benenti systems constitute the simplest, but still very wide, class
of all possible St\"{a}ckel separable systems. It can be shown \cite{M+A
JphysA2008} that this class contains all quadratic in momenta St\"{a}ckel
separable systems since all other systems of this type are constructed from
(\ref{BenSR}) by appropriate generalized St\"{a}ckel transforms and related
reciprocal transformations. Below we remind some established facts about
Benenti systems.

The relations (\ref{BenSR}) are linear in the coefficients $a_{i}$. Solving
these relations with respect to $a_{i}$ we obtain%

\begin{equation}
a_{i}=A\mu^{T}K_{i}G(f)\mu+BV_{i}(\gamma)\equiv H_{i}\text{, \ }%
i=1,\ldots,n\text{,}\label{HamB}%
\end{equation}
where we use the notation $\lambda=(\lambda_{1},\ldots,\lambda_{n})^{T}$ and
$\mu=(\mu_{1},\ldots,\mu_{n})^{T}$. Functions $H_{i}$ defined as the right
hand sides of the solution (\ref{HamB}) can be (locally) interpreted as $n$
quadratic in momenta $\mu$ Hamiltonians on the phase space $M=T^{\ast
}\mathcal{Q}$ cotangent to a Riemannian manifold $\mathcal{Q}$ equipped with
the contravariant metric tensor $G(f)$ depending on function $f $ only. As
mentioned above, these Hamiltonians are in involution with respect to the
canonical Poisson bracket on $T^{\ast}\mathcal{Q}$. Moreover, they are
separable in the sense of Hamilton-Jacobi theory since they by the very
definition satisfy St\"{a}ckel relations (\ref{SRalg}). The objects $K_{i}$ in
(\ref{HamB}) can be interpreted as $(1,1)$-type Killing tensors on
$\mathcal{Q}$ related to the family of metrics $G(f)$. The scalar functions
$V_{i}(\gamma)$ depend only on the function $\gamma$ and can be considered as
separable potentials. Further, the metric tensor $G$ and all the Killing
tensors $K_{i}$ \ are diagonal in $\lambda$-variables. More specifically, in
\thinspace$\lambda$-variables they attain the form%

\begin{equation}
G(f)=\operatorname*{diag}\left(  \frac{f(\lambda_{1})}{\Delta_{1}}%
,\ldots,\frac{f(\lambda_{n})}{\Delta_{n}}\right)  \text{ \ \ with }\Delta_{i}=%
{\textstyle\prod\limits_{j\neq i}}
(\lambda_{i}-\lambda_{j})\label{metryka}%
\end{equation}
and%
\[
K_{i}=-\operatorname*{diag}\left(  \frac{\partial q_{i}}{\partial\lambda_{1}%
},\cdots,\frac{\partial q_{i}}{\partial\lambda_{n}}\right)  \text{
\ \ \ }i=1,\ldots,n
\]
respectively. Here and below $q_{i}=q_{i}(\lambda)$ are Vi\`{e}te polynomials
(signed symmetric polynomials) in $\lambda$:%
\begin{equation}
q_{i}(\lambda)=(-1)^{i}%
{\displaystyle\sum\limits_{1\leq s_{1}<s_{2}<\ldots<s_{i}\leq n}}
\lambda_{s_{1}}\ldots\lambda_{s_{i}}\text{, \ \ }i=1,\ldots,n.\label{defq}%
\end{equation}
that can also be considered as new coordinates on \ the Riemannian manifold
$\mathcal{Q}$ (we will then refer to them as Vi\`{e}te coordinates). Notice
that the Killing tensors do not depend on a particular choice of $f$ and
$\gamma$.

\begin{remark}
The general $n-$time (simultaneous) solution for Hamilton equations
(\ref{2.2}) associated with all the Hamiltonians (\ref{HamB})\ attains the form
\end{remark}%

\begin{equation}
t_{i}+c_{i}=\pm\frac{1}{2\sqrt{A}}\sum_{r=1}^{n}%
{\displaystyle\int}
\frac{\lambda_{r}^{n-i}}{\sqrt{f(\lambda_{r})\left(
{\textstyle\sum\nolimits_{j=1}^{n}}
a_{j}\lambda_{r}^{n-j}-B\gamma(\lambda_{r})\right)  }}d\lambda_{r}\text{,
\ \ \ }i=1,\ldots,n\text{.}\label{gensol}%
\end{equation}
To see this it is enough to integrate the related Hamilton-Jacobi problem.
Now, with every Hamiltonian $H_{i}$ in (\ref{HamB}) we can associate the
corresponding inverse Legendre mapping (fiber derivartive) $\mathcal{L}%
_{i}^{-1}:T^{\ast}\mathcal{Q\rightarrow}T\mathcal{Q}$ given in the natural
coordinates $(\lambda,\mu)$ on $T^{\ast}\mathcal{Q}$ and $(\lambda
,\lambda_{t_{i}})$ on $T\mathcal{Q}$ respectively by
\[
\mathcal{L}_{i}^{-1}(\lambda,\mu)=(\lambda,2AK_{i}G\mu)=(\lambda
,\lambda_{t_{i}}).
\]
Performing $n$ corresponding Legendre transforms we obtain $n$ Lagrangians
$L_{i}:T\mathcal{Q\rightarrow}\mathbf{R}$ given explicitly by%
\begin{equation}
L_{i}(\lambda,\lambda_{t_{i}})=\frac{1}{4A}\lambda_{t_{i}}^{T}gK_{i}%
^{-1}\lambda_{t_{i}}-BV_{i}(\gamma),\label{Li}%
\end{equation}
where $g=G^{-1}$ is the corresponding covariant metric tensor. These
Lagrangians give rise to $n$ systems of Euler-Lagrange equations
\begin{equation}
E_{t_{i}}(L_{i})=0\text{, \ \ }i=1,\ldots,n\label{EL}%
\end{equation}
where $E_{t_{i}}$ is the Euler-Lagrange operator with respect to the
independent variable $t_{i}$. By construction, the solutions (\ref{gensol})
are also general solutions for all the Euler-Lagrange equations (\ref{EL}). It
means that for a particular $i$ the general solution of Euler-Lagrange
equation $E_{t_{i}}(L_{i})=0$ is given by (\ref{gensol}) where $t_{j}$ is
constant for $j\neq i$.

\section{Dispersionless Killing systems of Benenti type}

We have shown above that separation relations (\ref{BenSR}) lead to separable
Liouville systems (Benenti systems) and we also presented its general
solutions. It turns out that with the set of Killing tensors $K_{i}$ of any
Benenti system we can also relate a set of first order evolutionary PDE's.

\begin{definition}
\cite{MA} For any fixed $j\in\{1,...,n\}$, any of the following $n$ systems of
evolutionary PDE's of the form%
\begin{equation}
\lambda_{t_{i}}=K_{i}K_{j}^{-1}\lambda_{t_{j}}\equiv Z_{ij}(\lambda
,\lambda_{t_{j}})\text{, \ \ }i=1,\ldots,n.\label{Kilsys}%
\end{equation}
(where $\lambda=\left(  \lambda_{1},\ldots\lambda_{n}\right)  ^{T}$) with
$K_{i}$ being Killing tensors of a Benenti system (\ref{HamB}), will be called
\emph{a dispersionless Killing system of Benenti type}.
\end{definition}

The chosen variable $t_{j}$ in (\ref{Kilsys}) plays the role of a space
variable while the remaining variables $t_{i}$ should then be considered as
evolution parameters (times). Equations (\ref{Kilsys}) constitute a set of $n
$ integrable dispersionless equations, belonging to the class of so-called
\emph{weakly nonlinear} hydrodynamic-type systems, i.e. that are
semi-Hamiltonian in the sense of Tsarev \cite{tsar1},\cite{tsar2} and linearly
degenerate \cite{ferap1}, where the variables $\lambda_{i}$ are the Riemann
invariants for the system. For a chosen $j$, the systems (\ref{Kilsys}) can be
considered as $n$ vector fields $Z_{ij}$ (with a fixed $j$) on some infinite
dimensional function space $\mathcal{M}_{j}$ of functions $(\lambda_{1}%
(t_{j}),\ldots,\lambda_{n}(t_{j}))$. It can be shown that

\begin{proposition}
On $\mathcal{M}_{j}$ (i.e. for a fixed $j$) the vector fields $Z_{ij}$
commute:
\[
\left[  Z_{ij},Z_{kj}\right]  =0\ \ \ \ \ \ i,k=1,\ldots,n.
\]

\end{proposition}

The following proposition is crucial for our study:

\begin{proposition}
Every mutual solution $\lambda(t_{1},\ldots,t_{n})$ (given by (\ref{gensol}))
of all Hamiltonian systems (\ref{2.2}) with Hamiltonians of Benenti type
(\ref{HamB}) is also a solution of all corresponding Killing system in
(\ref{Kilsys}).
\end{proposition}

\begin{proof}
For the class of Benenti systems (\ref{HamB}) the spatial part of (\ref{2.2})
attains the form
\begin{equation}
\lambda_{t_{i}}=\frac{\partial H_{i}}{\partial\mu}=2AK_{i}G\mu
,\ \ \ \ \ \ \ \ i=1,...,n.\label{eliminuj}%
\end{equation}
So, for any fixed $j\in\{1,...,n\}$ we can eliminate the momenta $\mu$ from
(\ref{eliminuj}). This yields (\ref{Kilsys}).
\end{proof}

Thus, all the solutions (\ref{gensol}) are also solutions of all $n$ Killing
systems in (\ref{Kilsys}). Moreover, we have

\begin{theorem}
The $n$-time general solution of all the Killing systems in (\ref{Kilsys}) is
given by%
\begin{equation}
t_{i}+c_{i}=\sum_{r=1}^{n}%
{\displaystyle\int}
\frac{\lambda_{r}^{n-i}}{\varphi_{r}(\lambda_{r})}d\lambda_{r}%
,\ \ \ \ i=1,...,n\label{genKil}%
\end{equation}
(where $\varphi_{r}$ are arbitrary functions of one variable)
\end{theorem}

The proof of this statement can be found in \cite{ferap1}. Thus (\ref{genKil})
indeed encompasses all the solutions (\ref{gensol}). We also see that any
solution (\ref{genKil}) can be written in the form (\ref{gensol}) with
appropriately chosen $f$, $\gamma$, $a_{j},$ $A$ and $B$) so that on a given
surface of fixed values of all $a_{i}$ (for example for zero-energy solutions,
see below) with any such solution we can associate infinitely many
corresponding St\"{a}ckel systems (\ref{2.2}) sharing the same solution.

Consider now solutions (\ref{gensol}) as a specific class of solutions of
(\ref{Kilsys}). Since this class of solutions - by construction - satisfies
all the Euler-Lagrange equations (\ref{EL}) we can treat these equations as
additional bonds that these solutions satisfies. We can therefore use these
bonds to express some variables $\lambda_{i}$ by other $\lambda$s. Thus,
within the class (\ref{gensol}) of solutions of the Killing system
(\ref{Kilsys}) we can perform a variable elimination (reparametrization) that
turns (\ref{Kilsys}) into entirely new sets of evolutionary PDE's. \ Below we
demonstrate that in some carefully chosen cases this reparametrization turns
systems (\ref{Kilsys}) into systems with dispersion (soliton hierarchies) with
the solution (\ref{gensol}) being also a solution of these new systems with
dispersion. Specifically, we will produce this way all coupled KdV hierarchies
as well as new interesting classes of their solutions: finite-gap and rational
solutions. We will also demonstrate that our hierarchies indeed are the well
known cKdV hierarchies obtained by Antonowicz and Fordy through the
energy-dependent Schr\"{o}dinger spectral problem \cite{AF}.

From now on we will assume that%
\begin{equation}
f=\lambda^{m}\text{, \ \ }\gamma=\lambda^{k}\text{, \ \ \ }m,k\in
\mathbf{Z}\label{zal}%
\end{equation}
so that (\ref{BenSR}) attains the form%
\begin{equation}%
{\displaystyle\sum\limits_{j=1}^{n}}
a_{j}\lambda_{i}^{n-j}=A\lambda_{i}^{m}\mu_{i}^{2}+B\lambda_{i}^{k}\text{,
\ \ \ }i=1,\ldots,n.\label{BenSRspec}%
\end{equation}
We will denote the metric associated with $f=\lambda^{m}$ through
(\ref{metryka}) by $G^{(m)}$. It can be shown that for $m=0,\ldots,n$ the
metric $G^{(m)}$ is of zero curvature while the metric $G^{(n+1)}$ has
non-zero constant curvature. The separable potentials associated with
$\gamma=\lambda^{k}$ will be denoted by $V_{i}^{(k)}$. The family of separable
potentials $V_{i}^{(k)}$ can be constructed recursively \cite{maciej} by%

\begin{equation}
V_{i}^{(k+1)}=V_{i+1}^{(k)}-q_{i}V_{1}^{(k)}\text{ with }V_{i}^{(0)}%
=\delta_{in}\label{wprost}%
\end{equation}
where we put $V_{i}^{(k)}=0$ for $i<0$ or $i>n$. The first potentials are
trivial: $V_{i}^{(k)}=\delta_{i,n-k}$ for $k=0,1,\ldots,n-1$. The first
nontrivial potential is $V_{i}^{(n)}=-q_{i},$ For \thinspace$k>n$ the
potentials $V_{i}^{(k)}$ become complicated polynomial functions of $\lambda$.

The Lagrangians (\ref{Li}) for any specific choice of $m$ and $n~$\ are
denoted as $L_{i}^{n,m,k}$ so that $L_{i}^{n,m,k}=\frac{1}{4A}\lambda_{t_{i}%
}^{T}g^{(m)}K_{i}^{-1}\lambda_{t_{i}}-BV_{i}^{(k)}$.

From now on we will choose the representation $j=1$ in (\ref{Kilsys}) so that
the variable $t_{1}$ plays the role of the space variable. We denote therefore
this variable as $x$: $t_{1}=x$. The case of higher $j$ is not discussed in
this paper. Thus, we consider the Killing systems of the form%
\begin{equation}
\lambda_{t_{i}}=K_{i}\lambda_{x}\equiv Z_{i}^{n}(\lambda,\lambda_{x})\text{,
\ \ }i=1,\ldots,n.\label{Kilsys1}%
\end{equation}
where the new upper index $n$ in the $i$th vector field $Z_{i}$ denotes the
number of its components; note also that the second lower index in $Z$ is now
always $1$ and can therefore be omitted. Also, from now on we denote the
Lagrangian $L_{1}^{n,m,k}$ simply as $L^{n,m,k}$ (to avoid the unnecessary
index) so that%
\begin{equation}
L^{n,m,k}=\frac{1}{4A}\lambda_{x}^{T}g^{(m)}\lambda_{x}-BV_{1}^{(k)},\text{
\ \ }i=1,\ldots,n.\label{L1l}%
\end{equation}
In order to perform the aforementioned elimination procedure we will first
pass to Vi\`{e}te coordinates (\ref{defq}). The Killing systems (\ref{Kilsys1}%
) are tensorial so in Vi\`{e}te coordinates they have the form $q_{t_{r}%
}=K_{r}(q)q_{x}$ or, explicitly
\begin{equation}
\frac{d}{dt_{i}}q_{j}=(q_{j+i-1})_{x}+\sum_{k=1}^{j-1}\left(  q_{k}\left(
q_{j+i-k-1}\right)  _{x}-q_{j+i-k-1}\left(  q_{k}\right)  _{x}\right)
\equiv\left(  Z_{i}^{n}\left[  q\right]  \right)  ^{j}\text{ \ \ \ \ }%
i,j=1,\ldots,n\label{Kilq}%
\end{equation}
where $q_{\alpha}=0$ as soon as $\alpha>n$ and $\left(  Z_{i}^{n}\left[
q\right]  \right)  ^{j}$ denotes the $j$-th component of the vector field
$Z_{i}^{n}\left[  q\right]  \equiv Z_{i}^{n}(q,q_{x})$ (here and below the
symbol $f[q]$ will denote a differential function of $q$, that is a function
depending on $q$ and a finite number of its derivatives). One can see that
$\left(  Z_{i}^{n}\left[  q\right]  \right)  ^{j}=\left(  Z_{j}^{n}\left[
q\right]  \right)  ^{i}$ for all $i,j=1,\ldots,n$. Further, in Vi\`{e}te
coordinates the Lagrangian (\ref{L1l}) takes the form%
\begin{equation}
L^{n,m,k}=L^{n,m,k}(q,q_{x})=\frac{1}{4A}q_{x}^{T}g^{(m)}q_{x}-BV_{1}%
^{(k)},\text{ \ \ }i=1,\ldots,n,\label{L1}%
\end{equation}
where $g_{ij}^{(m)}=V_{1}^{(2n-m-i-j)}$ \cite{bensol}. The Euler-Lagrange
operator $E_{t_{1}}$in (\ref{EL}) will be simply denoted as $E,$ so in
Vi\`{e}te coordinates
\[
E=\left(  E_{1},\ldots E_{n}\right)  ,\text{ \ \ }E_{i}=\frac{\delta}{\delta
q_{i}}.
\]

\begin{theorem}
\label{symmetry}Lagrangian (\ref{L1}) satisfies the following symmetry relations:
\end{theorem}

\begin{enumerate}
\item for $\alpha=1,\ldots,n-1$%
\begin{equation}
E_{i}\left(  L^{n,m,k}\right)  =E_{i-\alpha}\left(  L^{n,m+\alpha,k-\alpha
}\right)  ,\ \ \ \ \ \ \ i=\alpha+1,...,n,\label{wtyl}%
\end{equation}

that can also be written as%
\begin{equation}
E_{i}\left(  L^{n,m,k}\right)  =E_{i+\alpha}\left(  L^{n,m-\alpha,k+\alpha
}\right)  ,\ \ \ \ \ \ \ i=1,...,n-\alpha.\label{wprzod}%
\end{equation}

\item
\begin{equation}
E_{l}\left(  L^{n,0,2n+\sigma}\right)  =E_{l+1}\left(  L^{n+1,0,2n+\sigma
+2}\right)  ,\ \ \ \sigma=1,...,n-1,\ \ \ l=\sigma+1,...,n.\label{19a}%
\end{equation}

\end{enumerate}

The proof of this theorem can be found in \cite{bensol}. \ This seemingly
technical theorem guarantees that the form of Euler-Lagrange equations
survives the passage from $n$-component to $(n+1)$-component Killing system
and hence it will be crucial for the construction of soliton hierarchies
below. The index $\sigma$ will be related with the number of components of the
obtained soliton systems.

For our further considerations we will also need a hierarchy of infinite
Killing systems%
\begin{equation}
\frac{d}{dt_{i}}q_{j}=(q_{j+i-1})_{x}+\sum_{k=1}^{j-1}\left(  q_{k}\left(
q_{j+i-k-1}\right)  _{x}-q_{j+i-k-1}\left(  q_{k}\right)  _{x}\right)
\equiv\left(  Z_{i}^{\infty}\left[  q\right]  \right)  ^{j}\text{
\ \ \ \ }i,j=1,\ldots\infty\label{Kilinf}%
\end{equation}
that is known as the universal hierarchy and has been considered in \cite{al1}.

\section{Coupled KdV hierarchies and their solutions}

We now shortly remind the reader our specific elimination procedure from
\cite{bensol} that turns the dispersionless Killing systems (\ref{Kilq}) into
cKdV hierarchies. More specifically, we show how to produce $s$ (with
$s\in\mathbf{N}$) $N$-component ($N\in\mathbf{N}$) commuting vector fields
(evolutionary systems) by eliminating some variables from a set of Killing
systems (\ref{Kilq}) with the help of Euler-Lagrange equations for an
appropriate Lagrangian $L^{n,m,k}$. The crucial for this procedure is that if
applied to $s+1$ instead of $s$ it yields the same set of $s$ commuting vector
fields plus an extra vector field that commutes with the first $s$ fields.
That means that this procedure leads in fact to an infinite hierarchy of
commuting vector fields in the sense that for arbitrary $s$ we can produce
first $s$ vector fields from the same infinite sequence of commuting vector
fields. Moreover it turns out that this way we produce vector fields with
dispersion (soliton systems), namely well-known coupled KdV hierarchies of
Antonowicz and Fordy \cite{AF} (in a different parametrization). Details are
as follows.

Firstly, we choose $A=1$ and $B=-1$. This specific choice of $A$ and $B$ is
introduced only for a smoother identification of our systems with the
aforementioned cKdV hierarchies; the elimination procedure works otherwise for
arbitrary values of $A$ and $B$. \ Consider all $N$ possible splittings
\[
N=\sigma+\alpha\text{ \ with }\sigma\in\left\{  1,\ldots,N\right\}  \text{
\ and \ }\alpha\in\left\{  0,\ldots,N-1\right\}  .
\]
Every such splitting leads to a separate hierarchy. Consider also the Killing
systems (\ref{Kilq}), written in a shorthand way as:%
\begin{equation}
q_{t_{r}}=Z_{r}^{n}\left[  q_{1,}\ldots,q_{n}\right]  \text{, \ }%
r=1,\ldots,n\label{Killingsigma}%
\end{equation}
where $q=\left(  q_{1},\ldots q_{n}\right)  ^{T}$.

\begin{remark}
\label{kompl}The first $s=n-N+1$ equations in (\ref{Killingsigma}) are such
that their first $N$ components coincide with the corresponding components of
the infinite Killing hierarchy (\ref{Kilinf}). The remaining $n-s$ equations
in (\ref{Killingsigma}) are incomplete with respect to the infinite hierarchy
(\ref{Kilinf}) since beginning with the flow $s+1$ systems (\ref{Kilinf})
contain at its first $N$ components also the variables $q_{n+1},\ldots
,q_{n+N-1}$.
\end{remark}

Let us now choose $m=-\alpha$ and $k=2n+N$ in (\ref{zal}) so that
$f=\lambda^{-\alpha}$ and $\gamma=\lambda^{2n+N}$ and consider \emph{the last}
$n-N$ Euler-Lagrange equations associated with $L^{n,-\alpha,2n+N}$. One can
show \cite{bensol} that they have the form:%
\begin{equation}%
\begin{array}
[c]{l}%
E_{N+1}\left(  L^{n,-\alpha,2n+N}\right)  \equiv2q_{n}+\varphi_{n-N+1}%
^{(\alpha)}[q_{1},...,q_{n-1}]=0,\\
E_{N+2}\left(  L^{n,-\alpha,2n+N}\right)  \equiv2q_{n-1}+\varphi
_{n-N}^{(\alpha)}[q_{1},...,q_{n-2}]=0,\\
\vdots\\
E_{n}\left(  L^{n,-\alpha,2n+N}\right)  \equiv2q_{N+1}+\varphi_{1}^{(\alpha
)}[q_{1},...,q_{N}]=0.
\end{array}
\label{EL1}%
\end{equation}
Due to their structure, equations (\ref{EL1}) can be explicitly solved with
respect to the variables $q_{N+1},\ldots,q_{n}$ which yields $q_{N+1}%
,\ldots,q_{n}$ as some differential functions of $q_{1},\ldots,q_{N} $:%

\begin{equation}%
\begin{array}
[c]{l}%
q_{N+1}=f_{1}^{(\alpha)}\left[  q_{1},\ldots,q_{N}\right] \\
\vdots\\
q_{n}=f_{n-N+1}^{(\alpha)}\left[  q_{1},\ldots,q_{N}\right]  .
\end{array}
\label{eli2}%
\end{equation}
Naturally, the solutions (\ref{gensol}) (with our choice of $f$ and $\gamma$)
solve both (\ref{Killingsigma}) and (\ref{EL1}). Thus, \emph{within the class
(\ref{gensol}) of solutions (\ref{genKil})}, we can use the Euler-Lagrange
equations (\ref{EL1}) or rather their solved form (\ref{eli2}) to successively
express (eliminate) the variables $q_{N+1},\ldots,q_{n}$ as differential
functions of $q_{1},\ldots,q_{N}$ in (\ref{Killingsigma}). Plugging
(\ref{eli2}) into (\ref{Killingsigma}) produces $n$ vector fields with
$N=\sigma+\alpha$ components:
\begin{equation}
\overline{q}_{t_{r}}=\overline{Z}_{r}^{n,N,\alpha}\left[  \overline{q}\right]
\text{\ \ \ \ }r=1,\ldots,n\text{, \ \ }\alpha\in\left\{  0,\ldots,N-1\right\}
\label{cel0}%
\end{equation}
(with $\overline{q}=\left(  q_{1},\ldots,q_{N}\right)  ^{T}$) . The higher
components of (\ref{Killingsigma}) disappear after this elimination within our
class (\ref{gensol}) of solutions. Moreover, since the first $s$ equations in
(\ref{Killingsigma}) are complete in the sense of Remark \ref{kompl} it can be
shown that%

\[
\overline{Z}_{r}^{n,N,\alpha}\left[  \overline{q}\right]  =\overline{Z}%
_{r}^{N,\alpha}\left[  \overline{q}\right]  \text{\ \ \ \ \ \ \ \ }%
r=1,\ldots,s\text{, }%
\]
meaning that the first $s=n-N+1$ equations in (\ref{cel0}) do not depend on
$n$. Observe also that equations (\ref{EL1}) do not depend on $n.$ Actually,
if $n$ increases to $n^{\prime}$ the last $n-N$ equations in (\ref{EL1}) with
this new $n^{\prime}$ will remain unaltered. This means that we can repeat
this elimination procedure by taking $n^{\prime}=n+1$ instead of $n$ (so that
$s$ increases to $s+1$ and $k=2n+N$ increases to $2(n+1)+N=k+2$ while $\sigma$
and $\alpha$ are kept unaltered. This new procedure (with $n^{\prime}=n+1$
instead of $n$) will therefore lead to a sequence of $s+1$ autonomous
$N=(\sigma+\alpha)$-component systems in which the first $s$ systems will
coincide with the corresponding systems obtained from the previous procedure
(with $s$). This way we can obtain arbitrary long sequences of the same
infinite set of commuting vector fields (soliton hierarchy):%

\begin{equation}
\overline{q}_{t_{r}}=\overline{Z}_{r}^{N,\alpha}\left[  \overline{q}\right]
\text{\ \ \ \ }r=1,2,\ldots\infty\text{, \ \ }\alpha\in\left\{  0,\ldots
,N-1\right\}  .\label{cel1}%
\end{equation}
The second index $\alpha$ in (\ref{cel1}) denotes different hierarchies. It
can be shown \cite{bensol} that vector fields $\overline{Z}_{r}^{N,\alpha}%
$commute%
\[
\left[  \overline{Z}_{i}^{N,\alpha},\overline{Z}_{j}^{N,\alpha}\right]
=0\text{ for any }i,j\in\mathbf{N.}%
\]
Note also that functions $f_{i}^{(\alpha)}$ in (\ref{eli2}) depend on $\alpha$
so that indeed the procedure leads to $N$ different hierarchies. Now, the $n$
functions $\lambda_{i}(t_{1},\ldots,t_{n})$ given implicitly by the system of
equations%
\begin{equation}
t_{i}+c_{i}=\pm\frac{1}{2}\sum_{r=1}^{n}%
{\displaystyle\int}
\frac{\lambda_{r}^{n-i+\alpha/2}}{\sqrt{\Delta_{r}^{N}}}d\lambda_{r}\text{
\ \ \ }i=1,\ldots,n\label{rb0}%
\end{equation}
with
\[
\Delta_{r}^{N}=\lambda_{r}^{2n+N}+%
{\textstyle\sum\nolimits_{j=1}^{n}}
a_{j}\lambda_{r}^{n-j}=%
{\textstyle\prod\nolimits_{i=1}^{2n+N}}
(\lambda_{r}-E_{i})
\]
are solutions of the first $n$ equations of the $N$-component hierarchy
(\ref{cel1}) with $N=\sigma+\alpha$. The reason is that equations (\ref{rb0})
are just equations (\ref{gensol}) with $f=\lambda^{-\alpha}$ and
$\gamma=\lambda^{2n+N}$ so they clearly solve all equations (\ref{cel0}).
Moreover, it can be shown, that these solutions are zero on $q_{n+1}%
,\ldots,q_{n+N-1}$ expressed as differential functions of $q_{1},\ldots,q_{N}$
through an appropriate system (\ref{EL1}) (with $n^{\prime}=n+N-1$). It means
that (\ref{rb0}) indeed solve the first $n$ equations in (\ref{cel1}).

Consider now the following infinite multi-Lagrangian "ladder" of
Euler-Lagrange equations of the form%
\begin{equation}
E_{1}\left(  L^{n,m+j-1,k-j+1}\right)  =E_{2}\left(  L^{n,m+j-2,k-j+2}\right)
=\cdots=E_{n}(L^{n,m+j-n,k-j+n})\label{ladder}%
\end{equation}
with fixed $m,k\in\mathbf{Z}$ and with $j=\ldots,-1,0,1,\ldots$ (the
multi-Lagrangian form of (\ref{ladder}) is due to Theorem \ref{symmetry}). The
equations (\ref{EL1}) that we use for variable elimination are then a part of
this infinite ladder with $m=-\alpha$ and $k=2n+\sigma+\alpha$ and with
$j=1,\ldots,n$. The equations (\ref{EL1}) are the only equations in the ladder
(\ref{ladder}) (for this specific choice of $m$ and $k$) that allow for the
elimination described above. All other equations are complicated polynomial
differential equations with no obvious structure that do not allow for any
elimination procedure. As a consequence, there exists more solutions of the
type (\ref{rb0}) associated with all the Lagrangians $L^{n,\beta
-\alpha,2N-\beta}$ for $\beta=1,\ldots,n-1.$ However, one can show that for
$\beta=2$ the variable $q_{n+N-1}$ is not zero on these solutions so that
these relations solve only first $n-1$ equations in (\ref{cel1}). More
generally, for any $\beta>1$ the obtained solutions $\lambda_{i}(t_{1}%
,\ldots,t_{n})$ will only satisfy first $n-\beta+1$ equations of the hierarchy
(\ref{cel1}). We can thus formulate the following theorem.

\begin{theorem}
\label{glowny}For any $\beta\in\left\{  0,\ldots,n-1\right\}  $ and any
$N=\sigma+\alpha<n$ the $n$ functions $\lambda_{i}(t_{1},\ldots,t_{n})$ given
implicitly by the system of equations%
\begin{equation}
t_{i}+c_{i}=\pm\frac{1}{2}\sum_{r=1}^{n}%
{\displaystyle\int}
\frac{\lambda_{r}^{n-i+\alpha/2-\beta/2}}{\sqrt{\Delta_{r}^{(N,\beta)}}%
}d\lambda_{r}\text{ \ \ \ }i=1,\ldots,n\label{rnia}%
\end{equation}
and with $\Delta_{r}^{(N,\beta)}$ given by
\[
\Delta_{r}^{(N,\beta)}=\lambda_{r}^{2n+N-\beta}+%
{\textstyle\sum\nolimits_{j=1}^{n}}
a_{j}\lambda_{r}^{n-j}=%
{\textstyle\prod\nolimits_{i=1}^{2n+N-\beta}}
(\lambda_{r}-E_{i})
\]
are solutions of the first $n-\beta+1$ (all $n$ for $\beta=0,1$) equations of
the $N$-component hierarchy (\ref{cel1}).
\end{theorem}

The variables $t_{1}=x,t_{2},\ldots,t_{n-\beta+1}$ in (\ref{rnia}) are
"dynamical times" (evolution parameters) of the hierarchy (\ref{cel1}) while
the variables $t_{n-\beta+2},\ldots,t_{n}$ are just free parameters (i.e. the
solutions (\ref{rnia}) do not solve flows higher than the flow number
$n-\beta+1$). Note also that, due to the structure of (\ref{defq}), all $n$
functions $\lambda_{i}(t_{1},\ldots,t_{n})$ obtained in (\ref{rnia}) are
necessary in order to compute $N$ functions $q_{i}(t_{1},\ldots,t_{n-\beta
+1})$ that solve (\ref{cel1}). In the case $N=1$ the solutions (\ref{rnia})
are finite-gap solutions for the KdV equation with the parameters $E_{i}$
playing the role of endpoints of forbidden zones. For $\beta>1$ these
solutions are up to our knowledge new. For $N>1$ all the solutions
(\ref{rnia}) are new.

\begin{remark}
\label{rownowazneKdV}For a fixed $\alpha\in\left\{  0,\ldots,N-1\right\}  $,
the following map%
\begin{align}
u_{r}  & =\frac{\partial V_{1}^{(N,2N)}}{\partial q_{N+1-r}}\text{,
\ }r=1,\ldots,N-\alpha\label{mapka}\\
u_{r}  & =E_{N+1-r}\left(  L^{N,N-\alpha,2N}\right)  \text{, \ \ }%
r=N-\alpha+1,\ldots,N\nonumber
\end{align}
(where $V_{1}^{(N,2N)}$ denotes the separable potential $V_{1}^{(2N)\text{ }}
$ in the dimension $N$) transforms the hierarchy (\ref{cel1}) to the hierarchy
generated by the spectral problem%
\begin{equation}
\left(  \lambda^{\alpha}\partial_{x}^{2}+%
{\textstyle\sum\nolimits_{i=1}^{N}}
u_{i}\lambda^{N-i}\right)  \Psi=\lambda^{N}\Psi.\label{sp}%
\end{equation}
This is the well known spectral problem of Antonowicz and Fordy leading to $N
$-component cKdV hierarchies, one for each $\alpha\in\left\{  0,\ldots
,N-1\right\}  $. Thus, $N$ hierarchies (\ref{cel1}) are nothing else as
(reparametrized) $N$ cKdV hierarchies from \cite{AF}.
\end{remark}

\section{Zero-energy solutions}

Let us now investigate - on few chosen examples - the nature of solutions
(\ref{rnia}) in the case that all $a_{i}$ vanish (zero-energy solutions). In
this case the solutions (\ref{rnia}) can easily be integrated yielding%
\begin{equation}
t_{i}+c_{i}=\pm\frac{1}{2-2i-\sigma}\sum_{r=1}^{n}\lambda_{r}^{1-i-\sigma
/2}\text{ \ \ \ }i=1,\ldots,n\label{zero}%
\end{equation}
and contain therefore no $\beta$ (are the same for all $\beta=0,\ldots,n-1 $)
and no $N$ except in $n=s+N-1$. Therefore, we obtain the following corollary:

\begin{corollary}
\label{wn} For any $N=\sigma+\alpha<n$, the $n$ functions $\lambda_{i}%
(t_{1},\ldots,t_{n})$ given implicitly by the system of equations (\ref{zero})
are solutions of the first $n$ equations of the $N$-component cKdV hierarchy
(\ref{cel1}).
\end{corollary}

Naturally, the solutions (\ref{zero}) also solve (on the surface $H_{i}=0$ for
all $i$) all St\"{a}ckel systems (\ref{HamB}) for which $f(\lambda
)\gamma(\lambda)=\lambda^{2n+\sigma}.$Observe also that the solutions
(\ref{zero}) solve all the Euler-Lagrange equations for the infinite sequence
of Lagrangians $L^{n,-\alpha+j,2n+N-j}$ where $j\in\mathbf{Z}$. As a
consequence, the second part of the map (\ref{mapka}) is zero on solutions
(\ref{zero}). The reason for this is that due to Theorem \ref{symmetry} the
expressions%
\[
E_{N+1-r}\left(  L^{N,N-\alpha,2N}\right)  \text{, \ \ }r=N-\alpha+1,\ldots,N
\]
can be written as%
\[
E_{r}\left(  L^{n,0,2n+\sigma}\right)  \text{, \ \ }r=n-\alpha+1,\ldots,n
\]
whereas all the above expressions are members of the sequence $L^{n,-\alpha
+j,2n+N-j}$, $j\in\mathbf{Z}$, so that they are zero on solutions
(\ref{zero}). This implies that as soon as $\alpha>0$ the solutions
(\ref{zero}) in the representation of Antonowicz and Fordy, reduce to the
solutions of the corresponding hierarchy with the same $\sigma$ but with
$\alpha=0$.

\textbf{Example 1.} We start with the case $N=1$. We wish to obtain the first
$s=3$ flows in (\ref{cel1}). There is now only one splitting $N=\sigma+\alpha$
possible, namely $\sigma=1$ and $\alpha=0$. This choice leads to the usual KdV
hierarchy. We have to take $n=s+N-1=3$. The Killing systems
(\ref{Killingsigma}) have in this case the form%
\begin{align}
\frac{d}{dt_{1}}\left(
\begin{array}
[c]{c}%
q_{1}\\
q_{2}\\
q_{3}%
\end{array}
\right)   &  =\left(
\begin{array}
[c]{c}%
q_{1,x}\\
q_{2,x}\\
q_{3,x}%
\end{array}
\right)  =Z_{1}^{3}\nonumber\\
\frac{d}{dt_{2}}\left(
\begin{array}
[c]{c}%
q_{1}\\
q_{2}\\
q_{3}%
\end{array}
\right)   &  =\left(
\begin{array}
[c]{c}%
q_{2,x}\\
q_{3,x}+q_{1}q_{2,x}-q_{2}q_{1,x}\\
q_{1}q_{3,x}-q_{3}q_{1,x}%
\end{array}
\right)  =Z_{2}^{3}\label{E1}\\
\frac{d}{dt_{3}}\left(
\begin{array}
[c]{c}%
q_{1}\\
q_{2}\\
q_{3}%
\end{array}
\right)   &  =\left(
\begin{array}
[c]{c}%
q_{3,x}\\
q_{1}q_{3,x}-q_{3}q_{1,x}\\
q_{2}q_{3,x}-q_{3}q_{2,x}%
\end{array}
\right)  =Z_{3}^{3}\nonumber
\end{align}
The Lagrangian $L^{n,-\alpha,2n+\sigma+\alpha}=L^{3,0,7}\,$\ is%

\[
L^{3,0,7\,}=\frac{1}{4}\left(  q_{1}^{2}-q_{2}\right)  q_{1,x}^{2}-\frac{1}%
{2}q_{1}q_{1,x}q_{2x}+\frac{1}{2}q_{1,x}q_{3,x}+\frac{1}{4}q_{2,x}^{2}%
+2q_{2}q_{3}-3q_{1}^{2}q_{3}-3q_{1}q_{2}^{2}+4q_{1}^{3}q_{2}-q_{1}^{5}%
\]
and the Euler-Lagrange equations (\ref{EL1}) attain the form%

\[%
\begin{array}
[c]{l}%
E_{2}\left(  L^{3,0,7}\right)  \equiv2q_{3}-6q_{1}q_{2}+4q_{1}^{3}+\frac{1}%
{4}q_{1,x}^{2}+\frac{1}{2}q_{1}q_{1,xx}-\frac{1}{2}q_{2,xx}=0,\\
E_{3}\left(  L^{3,0,7}\right)  \equiv2q_{2}-3q_{1}^{2}-\frac{1}{2}q_{1,xx}=0.
\end{array}
\]
Due to their structure, these equations can be solved with respect to
$q_{2},q_{3}$ yielding (\ref{eli2}) of the form%

\begin{equation}
q_{2}=\tfrac{1}{4}q_{1,xx}+\tfrac{3}{2}q_{1}^{2}\text{ \ \ , \ \ }q_{3}%
=\tfrac{1}{16}q_{1,xxxx}+\tfrac{5}{4}q_{1}q_{1,xx}+\tfrac{5}{8}q_{1,x}%
^{2}+\tfrac{5}{2}q_{1}^{3}.\label{eli2kdv}%
\end{equation}
Substituting (\ref{eli2kdv}) to the Killing systems (\ref{E1}) yields the
three one-component flows (\ref{cel1}):%
\begin{align}
q_{1,t_{1}}  & =q_{1,x}=\overline{Z}_{1}^{1,0}\nonumber\\
q_{1,t_{2}}  & =\frac{1}{4}q_{1,xxx}+3q_{1}q_{1,x}=\overline{Z}_{2}%
^{1,0}\label{poczkdv}\\
q_{1,t_{3}}  & =\frac{1}{16}q_{1,xxxxx}+\frac{5}{2}q_{1,x}q_{1,xx}+\frac{5}%
{4}q_{1}q_{1,xxx}+\frac{15}{2}q_{1}^{2}q_{1,x}=\overline{Z}_{3}^{1,0}\nonumber
\end{align}
which are just the first three flows of the KdV hierarchy. By taking larger
$s$ we can produce an arbitrary number of flows from the KdV hierarchy. Now,
according to Corollary \ref{wn}, the formula (\ref{zero}) yields some specific
solutions of all three flows in (\ref{poczkdv}). Explicitly, this formula
reads (with $x=t_{1},$ $c_{i}=0$ and $+$ sign in (\ref{zero})):%
\begin{align}
x  & =-%
{\textstyle\sum\nolimits_{i=1}^{3}}
z_{i}=-\rho_{1}\nonumber\\
t_{2}  & =-\frac{1}{3}%
{\textstyle\sum\nolimits_{i=1}^{3}}
z_{i}^{3}=-\frac{1}{3}\left(  \rho_{1}^{3}-3\rho_{1}\rho_{2}+3\rho_{3}\right)
\label{kaszana}\\
t_{3}  & =-\frac{1}{5}%
{\textstyle\sum\nolimits_{i=1}^{3}}
z_{i}^{5}=-\frac{1}{5}\left(  \rho_{1}^{5}-5(\rho_{1}\rho_{2}-\rho_{3}%
)(\rho_{1}^{2}-\rho_{2})\right) \nonumber
\end{align}
where $z_{i}=\lambda_{i}^{-1/2}$, $i=1,2,3$ and where $\rho_{1}=%
{\textstyle\sum\nolimits_{i=1}^{3}}
z_{i}$, $\rho_{2}=z_{1}z_{2}+z_{1}z_{3}+z_{2}z_{3}$ and $\rho_{3}=z_{1}%
z_{2}z_{3}$ are elementary symmetric polynomials in $z_{i}$. The right hand
sides of (\ref{kaszana}) follow from Newton formulas:%
\begin{equation}%
{\textstyle\sum\nolimits_{i=1}^{n}}
z_{i}^{m}=%
{\displaystyle\sum\limits_{\alpha_{1}+2\alpha_{2}+\ldots+n\alpha_{n}=m}}
(-1)^{a_{2}+\alpha_{4}+\alpha_{6}+\cdots}m\frac{\left(  \alpha_{1}+\alpha
_{2}+\cdots+\alpha_{n}-1\right)  !}{\alpha_{1}!\ldots\alpha_{n}!}\rho
_{1}^{\alpha_{1}}\rho_{2}^{\alpha_{2}}\ldots\rho_{n}^{\alpha_{n}}\text{,
\ \ }m=1,\ldots,n\label{Newton}%
\end{equation}
that can be easily extended to the case $m\geq n$ by taking larger $n$ and
putting all higher $\rho_{i}$ equal to zero. The system (\ref{kaszana}) can be
solved explicitly yielding the $3$-time solutions%
\begin{align}
\rho_{1}  & =-x\nonumber\\
\rho_{2}  & =\frac{15t_{3}+2x^{5}-15x^{2}t_{2}}{5\left(  x^{3}-3t_{2}\right)
}\label{Rrho}\\
\rho_{3}  & =\frac{-15t_{2}x^{3}-45t_{2}^{2}+x^{6}+45xt_{3}}{15\left(
x^{3}-3t_{2}\right)  }\nonumber
\end{align}
On the other hand, it is easy to see that%
\begin{equation}
q_{1}=\left(  \frac{\rho_{2}}{\rho_{3}}\right)  ^{2}-2\frac{\rho_{1}}{\rho
_{3}}\text{.}\label{tlumacz}%
\end{equation}
Plugging (\ref{Rrho}) into (\ref{tlumacz}) we finally obtain a $3$-time
solution of the first three flows (\ref{poczkdv}) of the KdV hierarchy. It has
a rather complicated, rational form:%
\begin{equation}
q_{1}(x,t_{2},t_{3})=\frac{-3\left(  675t_{3}^{2}-270t_{3}x^{5}+2x^{10}%
+675x^{4}t_{2}^{2}-1350xt_{2}^{3}\right)  }{\left(  -15t_{2}x^{3}-45t_{2}%
^{2}+x^{6}+45xt_{3}\right)  ^{2}},\label{rozwKdV}%
\end{equation}
By taking larger $s$ we can in this way obtain $s$-time solutions of first $s
$ flows of the KdV hierarchy. Finally, the map (\ref{mapka}) is in this case
trivial and reads $u_{1}=2q_{1}$.

\textbf{Example 2.} Let us consider the two field case: $N=2$. There are now
two splittings possible: $N=\sigma+\alpha=2+0$ and $N=\sigma+\alpha=1+1$. We
consider only the first two flows $s=2$ (i.e. only the first nontrivial flow)
so that $n=s+N-1=3$ as before and therefore the original Killing systems are
as before, i.e (\ref{E1}) - we just consider the first two of them. Let us
first consider the splitting $N=\sigma+\alpha=2+0$. The Lagrangian
$L^{n,-\alpha,2n+\sigma+\alpha}=L^{3,0,8}\,$\ is%

\begin{align*}
L^{3,0,8\,}  & =\frac{1}{4}\left(  q_{1}^{2}-q_{2}\right)  q_{1,x}^{2}%
-\frac{1}{2}q_{1}q_{1,x}q_{2x}+\frac{1}{2}q_{1,x}q_{3,x}+\frac{1}{4}%
q_{2,x}^{2}+\\
& +q_{3}^{2}-6q_{1}q_{2}q_{3}+4q_{3}q_{1}^{3}-q_{2}^{3}+6q_{2}^{2}q_{1}%
^{2}-5q_{2}q_{1}^{4}+q_{1}^{6}.
\end{align*}
(note that its kinetic energy part is the same as in $L^{3,0,7}$ above). The
formulas (\ref{EL1}) contain only one equation that can be solved with respect
to $q_{3}$ yielding (\ref{eli2}) of the form%

\[
q_{3}=3q_{1}q_{2}-2q_{1}^{3}+\frac{1}{4}q_{1,xx}.
\]
Substituting this to (\ref{Killingsigma}) (with $s=2$ now) yields the first
two flows of the first (i.e with $\alpha=0$) $2$-component cKdV hierarchy:%

\begin{equation}%
\begin{array}
[c]{c}%
\frac{d}{dt_{1}}\left(
\begin{array}
[c]{c}%
q_{1}\\
q_{2}%
\end{array}
\right)  =\left(
\begin{array}
[c]{c}%
q_{1,x}\\
q_{2,x}%
\end{array}
\right)  =\overline{Z}_{1}^{3,0}\\
\frac{d}{dt_{2}}\left(
\begin{array}
[c]{c}%
q_{1}\\
q_{2}%
\end{array}
\right)  =\left(
\begin{array}
[c]{c}%
q_{2,x}\\
2q_{2}q_{1,x}+4q_{1}q_{2,x}-6q_{1}^{2}q_{1,x}+\frac{1}{4}q_{1,xxx}%
\end{array}
\right)  =\overline{Z}_{2}^{3,0}%
\end{array}
\label{2ckdv}%
\end{equation}
The zero-energy solutions (\ref{zero}) (again with all $c_{i}=0$ and with the
plus sign only) attain now the form:%
\begin{align}
x  & =-\frac{1}{2}%
{\textstyle\sum\nolimits_{i=1}^{3}}
z_{i}=-\rho_{1}\nonumber\\
t_{2}  & =-\frac{1}{4}%
{\textstyle\sum\nolimits_{i=1}^{3}}
z_{i}^{2}=-\frac{1}{3}\left(  \rho_{1}^{3}-3\rho_{1}\rho_{2}+3\rho_{3}\right)
\label{kaszana2}\\
t_{3}  & =-\frac{1}{6}%
{\textstyle\sum\nolimits_{i=1}^{3}}
z_{i}^{3}=-\frac{1}{5}\left(  \rho_{1}^{5}-5(\rho_{1}\rho_{2}-\rho_{3}%
)(\rho_{1}^{2}-\rho_{2})\right) \nonumber
\end{align}
where $\rho_{i}$ again denote elementary symmetric polynomials in $z_{i}$ but
where now $z_{i}=\lambda_{i}^{-1}$. Solving (\ref{kaszana2}) yields the
following $3$-time solutions:%
\begin{align}
\rho_{1}  & =-2x\nonumber\\
\rho_{2}  & =2x^{2}+2t_{2}\label{Rrho2}\\
\rho_{3}  & =-\frac{4}{3}x^{3}-4xt_{2}-2t_{3}\nonumber
\end{align}
where the variable $t_{3}$ plays the role of a free parameter for
eqs.(\ref{2ckdv}). Moreover, we have%
\begin{equation}
q_{1}=-\frac{\rho_{2}}{\rho_{3}}\text{, \ }q_{2}=\frac{\rho_{1}}{\rho_{3}%
}.\label{tlumacz2}%
\end{equation}
Plugging (\ref{Rrho2}) into (\ref{tlumacz2}) we obtain the following solutions
for (\ref{2ckdv})%
\begin{equation}
q_{1}(x,t_{2},t_{3})=\frac{3(t_{2}+x^{2})}{3t_{3}+2x^{3}+6xt_{2}}\text{,
\ \ \ \ \ }q_{2}(x,t_{2},t_{3})=\frac{3x}{3t_{3}+2x^{3}+6xt_{2}}%
\label{rozw2kdv}%
\end{equation}
Note that Corollary \ref{wn} implies that the functions (\ref{rozw2kdv}) solve
first $n=3$ flows of the hierarchy (\ref{cel1}) with $N=2$ and $\alpha=0$. In
order to compute this third flow we just need to take $s=3$ in the elimination
procedure. The result is%
\[
\frac{d}{dt_{3}}\left(
\begin{array}
[c]{c}%
q_{1}\\
q_{2}%
\end{array}
\right)  =\left(
\begin{array}
[c]{c}%
3q_{2}q_{1,x}+3q_{1}q_{2,x}-6q_{1}^{2}q_{1,x}+\frac{1}{4}q_{1,xxx}\\
6q_{1}q_{2}q_{1,x}-18q_{1}^{3}q_{1,x}+6q_{1}^{2}q_{2,x}+\frac{3}{4}%
q_{1}q_{1,xxx}+3q_{2}q_{2,x}+\frac{1}{4}q_{2,xxx}%
\end{array}
\right)  =\overline{Z}_{3}^{3,0}.
\]
Let us finally pass to Antonowicz-Fordy variables. The map (\ref{mapka})
attains the form%
\[
u_{1}=2q_{1}\text{, \ \ \ }u_{2}=2q_{2}-3q_{1}^{2}%
\]
and it transforms both systems in (\ref{2ckdv}) to the representation of
Antonowicz and Fordy. Explicitly:%

\begin{equation}%
\begin{array}
[c]{c}%
\frac{d}{dt_{1}}\left(
\begin{array}
[c]{c}%
u_{1}\\
u_{2}%
\end{array}
\right)  =\left(
\begin{array}
[c]{c}%
u_{1,x}\\
u_{2,x}%
\end{array}
\right)  =\overline{Z}_{1}^{3,0}[u]\\
\frac{d}{dt_{2}}\left(
\begin{array}
[c]{c}%
u_{1}\\
u_{2}%
\end{array}
\right)  =\left(
\begin{array}
[c]{c}%
u_{2,x}+\frac{3}{2}u_{1}u_{1,x}\\
u_{2}u_{1,x}+\frac{1}{2}u_{1}u_{2,x}+\frac{1}{4}u_{1,xxx}%
\end{array}
\right)  =\overline{Z}_{2}^{3,0}[u]
\end{array}
\label{wzmAF}%
\end{equation}
In the $u$-variables the solutions (\ref{rozw2kdv}) yield solutions for
(\ref{wzmAF}) and attain the form%
\[
u_{1}(x,t_{2},t_{3})=\frac{6(t_{2}+x^{2})}{3t_{3}+2x^{3}+6xt_{2}}\text{,
\ \ \ \ \ }u_{2}(x,t_{2},t_{3})=\frac{3(6xt_{3}-5x^{4}-6x^{2}t_{2}-9t_{2}%
^{2})}{\left(  3t_{3}+2x^{3}+6xt_{2}\right)  ^{2}}.
\]

\textbf{Example 3.} Let us now consider the case $N=\sigma+\alpha=1+1$. Again,
we look for the first $s=2$ flows in (\ref{cel1}) with $N=2,$ $a=1$. We have
to take $n=s+N-1=3$ and the rather lenghty Lagrangian%

\begin{align*}
L^{n,-\alpha,2n+\sigma+\alpha}  & =L^{3,-1,8}=-\frac{1}{4}\left(  q_{1}%
^{3}-2q_{1}q_{2}+q_{3}\right)  q_{1,x}^{2}+\frac{1}{2}\left(  q_{1}^{2}%
-q_{2}\right)  q_{1,x}q_{2,x}\\
& -\frac{1}{2}q_{1}q_{1,x}q_{3,x}-\frac{1}{4}q_{1}q_{2,x}^{2}+\frac{1}%
{2}q_{2,x}q_{3,x}\\
& +q_{3}^{2}-6q_{1}q_{2}q_{3}+4q_{3}q_{1}^{3}-q_{2}^{3}+6q_{2}^{2}q_{1}%
^{2}-5q_{2}q_{1}^{4}+q_{1}^{6}%
\end{align*}
(the potential part is of course the same as in $L^{3,0,8}$ above). The
elimination equations (\ref{EL1}) yield again only one equation that solved
with respect to $q_{3}$ reads%
\[
q_{3}=-\frac{1}{8}q_{1,x}^{2}-2q_{1}^{3}+3q_{1}q_{2}-\frac{1}{4}q_{1}%
q_{1,xx}+\frac{1}{4}q_{2,xx}.
\]
Plugging this into two first flows in (\ref{E1}) we obtain the first two flows
of the second (i.e. with $\alpha=1$) $2$-field cKdV hierarchy:%
\begin{equation}%
\begin{array}
[c]{c}%
\frac{d}{dt_{1}}\left(
\begin{array}
[c]{c}%
q_{1}\\
q_{2}%
\end{array}
\right)  =\left(
\begin{array}
[c]{c}%
q_{1,x}\\
q_{2,x}%
\end{array}
\right)  =\overline{Z}_{1}^{3,1}\\
\frac{d}{dt_{2}}\left(
\begin{array}
[c]{c}%
q_{1}\\
q_{2}%
\end{array}
\right)  =\left(
\begin{array}
[c]{c}%
q_{2,x}\\
2q_{2}q_{1,x}+4q_{1}q_{2,x}-\frac{1}{2}q_{1,x}q_{1,xx}-6q_{1}^{2}q_{1,x}%
-\frac{1}{4}q_{1}q_{1,xxx}+\frac{1}{4}q_{2,xxx}%
\end{array}
\right)  =\overline{Z}_{2}^{3,1}%
\end{array}
\label{2ckdv2}%
\end{equation}
Now, the solutions (\ref{zero}) in this case attain precisely the form
(\ref{kaszana}), or (\ref{Rrho}) in solved form, since both $n$ and $\sigma$
are the same in both cases. However, (\ref{2ckdv2}) are two-component, so in
this case we need to express both $q_{1}$ and $q_{2}$ as functions of
$\rho_{i}:$%

\begin{equation}
q_{1}=\left(  \frac{\rho_{2}}{\rho_{3}}\right)  ^{2}-2\frac{\rho_{1}}{\rho
_{3}}\text{, \ }q_{2}=\left(  \frac{\rho_{1}}{\rho_{3}}\right)  ^{2}%
-2\frac{\rho_{2}}{\rho_{3}^{2}}.\label{tlumacz2.2}%
\end{equation}
Substituting (\ref{Rrho}) into (\ref{tlumacz2.2}) we finally arrive at a
$3$-time solution of (\ref{2ckdv2}) with $t_{3}$ as a free parameter:%

\begin{align}
q_{1}(x,t_{2},t_{3})  & =\frac{-3\left(  675t_{3}^{2}-270t_{3}x^{5}%
+2x^{10}+675x^{4}t_{2}^{2}-1350xt_{2}^{3}\right)  }{\left(  -15t_{2}%
x^{3}-45t_{2}^{2}+x^{6}+45xt_{3}\right)  ^{2}},\label{rozw2kdv2}\\
q_{2}(x,t_{2},t_{3})  & =\frac{45(x^{3}-3t_{2})(x^{5}+15x^{2}t_{2}-30t_{3}%
)}{\left(  -15t_{2}x^{3}-45t_{2}^{2}+x^{6}+45xt_{3}\right)  ^{2}}.\nonumber
\end{align}
As in the previous example, the above solutions solve first $n=3$ flows of
this cKdV hierarchy which means that they also solve the next flow in the
hierarchy (with the dynamical time $t_{3}$). We will however not write it
here. Finally, let us pass to the Antonowicz-Fordy representation.The map
(\ref{mapka}) is now%
\[
u_{1}=2q_{1}\text{, \ \ \ }u_{2}=2q_{2}-3q_{1}^{2}-\frac{1}{2}q_{1,xx}%
\]
so that (\ref{2ckdv2}) in Antonowicz-Fordy variables reads%

\begin{equation}%
\begin{array}
[c]{c}%
\frac{d}{dt_{1}}\left(
\begin{array}
[c]{c}%
u_{1}\\
u_{2}%
\end{array}
\right)  =\left(
\begin{array}
[c]{c}%
u_{1,x}\\
u_{2,x}%
\end{array}
\right)  =\overline{Z}_{1}^{3,1}[u]\\
\frac{d}{dt_{2}}\left(
\begin{array}
[c]{c}%
u_{1}\\
u_{2}%
\end{array}
\right)  =\left(
\begin{array}
[c]{c}%
u_{2,x}+\frac{3}{2}u_{1}u_{1,x}+\frac{1}{4}u_{1,xxx}\\
u_{2}u_{1,x}+\frac{1}{2}u_{1}u_{2,x}%
\end{array}
\right)  =\overline{Z}_{2}^{3,1}[u]
\end{array}
\label{wzmAF2}%
\end{equation}
In the $u$-variables the solutions (\ref{rozw2kdv2}) yield solutions for
(\ref{wzmAF2}) and attain the form%
\[
u_{1}(x,t_{2},t_{3})=\frac{-6\left(  675t_{3}^{2}-270t_{3}x^{5}+2x^{10}%
+675x^{4}t_{2}^{2}-1350xt_{2}^{3}\right)  }{\left(  -15t_{2}x^{3}-45t_{2}%
^{2}+x^{6}+45xt_{3}\right)  ^{2}}\text{, \ \ \ \ }u_{2}(x,t_{2},t_{3})=0
\]
which is nothing else than the solution (\ref{rozwKdV}) of the first three
flows of the KdV hierarchy (with $N=1$), in accordance with the observation at
the beginning of this section.

\begin{remark}
It can be shown that our method yields rational solutions only for $\sigma=1 $
and $\sigma=2$. For $\sigma>2$ our method leads to new \emph{implicit}
solutions of our cKdV hierarchies .
\end{remark}

\textbf{Example 4.} Let us thus finally investigate the case $N=3=\sigma
+\alpha=3+0$ that will lead to implicit solutions. We take again $s=2$ so that
$n=s+N-1=4$. The first $s=2$ Killing systems in (\ref{Killingsigma}) have now
the form%
\begin{align}
\frac{d}{dt_{1}}\left(
\begin{array}
[c]{c}%
q_{1}\\
q_{2}\\
q_{3}\\
q_{4}%
\end{array}
\right)   &  =\left(
\begin{array}
[c]{c}%
q_{1,x}\\
q_{2,x}\\
q_{3,x}\\
q_{4,x}%
\end{array}
\right)  =Z_{1}^{4}\label{K4}\\
\frac{d}{dt_{2}}\left(
\begin{array}
[c]{c}%
q_{1}\\
q_{2}\\
q_{3}\\
q_{4}%
\end{array}
\right)   &  =\left(
\begin{array}
[c]{c}%
q_{2,x}\\
q_{3,x}+q_{1}q_{2,x}-q_{2}q_{1,x}\\
q_{4,x}+q_{1}q_{3,x}-q_{3}q_{1,x}\\
q_{1}q_{4,x}-q_{4}q_{1,x}%
\end{array}
\right)  =Z_{2}^{4}\nonumber
\end{align}
The Lagrangian $L^{n,-\alpha,2n+N}=L^{4,0,11}$ yields one elimination equation
(\ref{EL1}):%
\[
E_{4}(L^{4,0,11})=2q_{4}+12q_{2}q_{1}^{2}-6q_{1}q_{3}-3q_{2}^{2}-5q_{1}%
^{4}-\frac{1}{2}q_{1,xx}=0.
\]
Solving this with respect to $q_{4}$ we obtain%
\[
q_{4}=-6q_{2}q_{1}^{2}+3q_{1}q_{3}+\frac{3}{2}q_{2}^{2}-\frac{5}{2}q_{1}%
^{4}+\frac{1}{4}q_{1,xx}.
\]
Substituting this into (\ref{K4}) yields two first flows of the 3-component
cKdV hierarchy (\ref{cel1}) with $\alpha=0$.%
\begin{align}
\frac{d}{dt_{1}}\left(
\begin{array}
[c]{c}%
\begin{array}
[c]{c}%
q_{1}\\
q_{2}%
\end{array}
\\
q_{3}%
\end{array}
\right)   &  =\left(
\begin{array}
[c]{c}%
q_{1,x}\\
q_{2,x}\\
q_{3,x}%
\end{array}
\right)  =\overline{Z}_{1}^{3,0}\label{ost}\\
\frac{d}{dt_{2}}\left(
\begin{array}
[c]{c}%
q_{1}\\
q_{2}\\
q_{3}%
\end{array}
\right)   &  =\left(
\begin{array}
[c]{c}%
q_{2,x}\\
q_{3,x}+q_{1}q_{2,x}-q_{2}q_{1,x}\\
2q_{3}q_{1,x}+4q_{1}q_{3,x}-6q_{1}^{2}q_{2,x}-12q_{1}q_{2}q_{1,x}%
+3q_{2}q_{2,x}+10q_{1}^{3}q_{1,x}+\frac{1}{4}q_{1,xxx}%
\end{array}
\right)  =\overline{Z}_{2}^{3,0}.\nonumber
\end{align}
The solutions (\ref{zero}) are now (again with all $c_{i}=0$ and with the plus
sign only):%
\begin{align}
x  & =-\frac{1}{3}%
{\textstyle\sum\nolimits_{i=1}^{4}}
z_{i}^{3}=-\frac{1}{3}\left(  \rho_{1}^{3}-3\rho_{1}\rho_{2}+3\rho_{3}\right)
\nonumber\\
t_{2}  & =-\frac{1}{5}%
{\textstyle\sum\nolimits_{i=1}^{4}}
z_{i}^{5}=-\frac{1}{5}\left(  \rho_{1}^{5}-5(\rho_{1}\rho_{2}-\rho_{3}%
)(\rho_{1}^{2}-\rho_{2})\right) \label{kaszana3}\\
t_{3}  & =-\frac{1}{7}%
{\textstyle\sum\nolimits_{i=1}^{4}}
z_{i}^{7}=-\frac{1}{7}\left(  \rho_{1}^{7}-7(\rho_{1}\rho_{2}-\rho_{3})\left(
(\rho_{1}^{2}-\rho_{2})^{2}+\rho_{1}\rho_{3}\right)  -7\rho_{4}(\rho_{1}%
^{3}-2\rho_{1}\rho_{2}+\rho_{3})\right) \nonumber\\
t_{4}  & =-\frac{1}{9}%
{\textstyle\sum\nolimits_{i=1}^{4}}
z_{i}^{9}=-\frac{1}{9}P_{9}\nonumber
\end{align}
with $z_{i}=\lambda_{i}^{-1/2}$ where $P_{9}$ is a complicated polynomial of
degree $9$ in $\rho_{i}$ that can be obtained from Newton formulas
(\ref{Newton}). This system can not be algebraically solved with respect to
$\rho_{i}$. However, if we embed the system (\ref{kaszana3}) in the system%
\begin{align}
\alpha & =-%
{\textstyle\sum\nolimits_{i=1}^{5}}
z_{i}=-\rho_{1}\nonumber\\
x  & =-\frac{1}{3}%
{\textstyle\sum\nolimits_{i=1}^{5}}
z_{i}^{3}=-\frac{1}{3}\left(  \rho_{1}^{3}-3\rho_{1}\rho_{2}+3\rho_{3}\right)
\nonumber\\
t_{2}  & =-\frac{1}{5}%
{\textstyle\sum\nolimits_{i=1}^{5}}
z_{i}^{5}=-\frac{1}{5}\left(  \rho_{1}^{5}-5(\rho_{1}\rho_{2}-\rho_{3}%
)(\rho_{1}^{2}-\rho_{2})\right)  +5\rho_{1}\rho_{4}+5\rho_{5}\label{totkasz}\\
t_{3}  & =-\frac{1}{7}%
{\textstyle\sum\nolimits_{i=1}^{5}}
z_{i}^{7}=-\frac{1}{7}\left(  \rho_{1}^{7}-7(\rho_{1}\rho_{2}-\rho_{3})\left(
(\rho_{1}^{2}-\rho_{2})^{2}+\rho_{1}\rho_{3}\right)  -7\rho_{4}(\rho_{1}%
^{3}-2\rho_{1}\rho_{2}+\rho_{3})\right) \nonumber\\
& +7\rho_{1}^{2}\rho_{5}\nonumber\\
t_{4}  & =-\frac{1}{9}%
{\textstyle\sum\nolimits_{i=1}^{5}}
z_{i}^{9}=-\frac{1}{9}Q_{9}\nonumber
\end{align}
where $\alpha$ is a parameter, and where $Q_{9}$ is a polynomial of degree $9$
in $\rho_{1},\ldots,\rho_{5}$ such that $\left.  Q_{9}\right\vert _{\rho
_{5}=0}=P_{9}$, then obviously the solution of (\ref{totkasz}) with the
condition $\rho_{5}=0$ will yield the solution for (\ref{kaszana3}). The
system (\ref{totkasz}) is algebraically solvable and yields%
\begin{equation}
\rho_{i}=R_{i}(\alpha,x,t_{2},t_{3},t_{4})\label{rhor}%
\end{equation}
where $R_{i}$ are complicated rational functions of their arguments. The
polynomial equation $\rho_{5}=0$ yields then implicitly a (multivalued)
function $\alpha=f(x,t_{2},t_{3},t_{4})$. Now, using the fact that
\[
q_{1}=-\left(  \frac{\rho_{3}}{\rho_{4}}\right)  ^{2}+2\frac{\rho_{2}}%
{\rho_{4}}\text{, \ }q_{2}=\left(  \frac{\rho_{2}}{\rho_{4}}\right)
-2\frac{\rho_{1}\rho_{3}}{\rho_{4}^{2}}+\frac{2}{\rho_{4}}\text{, \ }%
q_{3}=-\frac{(\rho_{1}^{4}+2\rho_{2}^{2}-4\rho_{1}^{2}\rho_{2}+4\rho_{1}%
\rho_{3}-4\rho_{4})}{\rho_{4}^{2}}%
\]
we arrive at an implicit solution of (\ref{ost}) of the form%
\[
\rho_{5}(\alpha,x,t_{2},t_{3},t_{4})=0,\ \ \ \ \ \ q_{i}=r_{i}\left(
\alpha,x,t_{2},t_{3},t_{4}\right)  \text{, \ \ }i=1,2,3,
\]
that is thus determined up to the implicitly expressed function $\alpha
=f(x,t_{2},t_{3},t_{4})$. The concrete formulas have been obtained with the
help of Maple and are too complicated to present them here. The map
(\ref{mapka}) does not simplify these solutions.

\section{Conclusions}

In this article we presented a method of constructing coupled Korteweg - de
Vries hierarchies from Benenti class of St\"{a}ckel separable systems. Our
method allows for producing certain class of solutions of these hierarchies
from solutions of corresponding St\"{a}ckel systems (Theorem \ref{glowny} and
Corollary \ref{wn}). For $N=1$ and $\beta=0,1$ these solutions are known but
for larger $\beta$ and for larger $N$ they all seem to be new. Also, for
$\sigma>2$ all our solutions are implicit in the sense described in Example 4
above. It has to be stressed that our method is general in the sense that
other separation relations lead to other hierarchies like for example coupled
Harry-Dym hierarchies. These possibilities will be studied in a separate paper.

Considering the class of solutions constructed in the paper one arrives at
natural question whether these solutins are somehow related with some class of
symmetry reductions. Some hint is given in the case $N=1$: the finite gap
solutions (\ref{rnia}) for $\beta=0$ and $\beta=1$ for the KdV can be obtained
from its stationary flows constructed with the help of first two local
Hamiltonian representations of the KdV hierarchy. This question is beyond the
scope of this paper but it certainly deserves a separate study.

\medskip

{\Large Acknowledgements}

\medskip

Both authors were partially supported by Swedish Research Council grant no VR
2006-7359 and by MNiSW research grant N202 4049 33.


\begin{thebibliography}{99}                                                                                               %
\bibitem {bensol}B\l aszak, M.; Marciniak, K., \emph{From St\"{a}ckel systems
to integrable hierarchies of PDE's: Benenti class of separation relations, }J.
Math. Phys. \textbf{47 }(2006) 032904\emph{\ }

\bibitem {AF}Antonowicz, M.; Fordy, A. P., \emph{Coupled KdV equations with
multi-Hamiltonian structures.} Phys. D \textbf{28} (1987), no. 3, 345--357.

\bibitem {bogo}Bogojavlenskii, O. I.; Novikov, S. P. \emph{The connection
between the Hamiltonian formalisms of stationary and nonstationary problems,}
Functional Anal. Appl. \textbf{10} (1976), 8--11.

\bibitem {antforstef}Antonowicz, M.; Fordy, A. P.; Wojciechowski, S.
\emph{Integrable stationary flows: Miura maps and bi-Hamiltonian structures.}
Phys. Lett. A \textbf{124} (1987) 143--150.

\bibitem {stefmar}Antonowicz, M.; Rauch-Wojciechowski, S. \emph{Restricted
flows of soliton hierarchies: coupled KdV and Harry Dym case.} J. Phys. A
\textbf{24} (1991), no. 21, 5043--5061.

\bibitem {skm}Rauch-Wojciechowski, S.; Marciniak, K.; Blaszak, M. \emph{Two
Newton decompositions of stationary flows of KdV and Harry Dym hierarchies}.
Phys. A \textbf{233} (1996), no. 1-2, 307--330.

\bibitem {cKdV}Marciniak, K., \emph{Coupled Korteweg-de Vries hierarchy with
sources and its Newton decomposition}. J. Math. Phys. \textbf{38} (1997), no.
11, 5739--5755.

\bibitem {mb1}B\l aszak, M. \emph{Multi-Hamiltonian theory of dynamical
systems.} Texts and Monographs in Physics. Springer-Verlag, Berlin, 1998.

\bibitem {al1}Alonso L.M.; Shabat A. B., \emph{Energy-dependent potentials
revisited: a universal hierarchy of hydrodynamic type}, Phys. Lett. \textbf{A
300} (2002) 58--64.

\bibitem {Sklyanin}Sklyanin, E. K. \emph{Separation of variables---new
trends.} Progr. Theoret. Phys. Suppl. \textbf{118}, (1995), 35--60.

\bibitem {M+A JphysA2008}Sergyeyev, A.; B\l aszak, M. \emph{Generalized
St\"{a}ckel transform and reciprocal transformations for finite-dimensional
integrable systems, }J. Phys. A: Math. Theor. \textbf{41} (2008) 105205

\bibitem {MA}B\l aszak, M.; Sergyeyev, A. \emph{Geometric construction for a
class of integrable weakly nonlinear hydrodynamic-type systems,} preprint
arXiv 0803.0308

\bibitem {tsar1}Tsarev, S. P. \emph{Poisson brackets and one-dimensional
Hamiltonian systems of hydrodynamic type.} Soviet Math. Dokl. \textbf{31}
(1985) 488--491.

\bibitem {tsar2}Tsarev, S. P. \ \emph{The geometry of Hamiltonian systems of
hydrodynamic type. The generalized hodograph method.} Math. USSR-Izv.
\textbf{37} (1991), no. 2, 397--419.

\bibitem {ferap1}Ferapontov, E. V. \emph{Integration of weakly nonlinear
hydrodynamic ssystems in Riemman invariants. }Phys. Lett. \textbf{A} (1991) 112-118

\bibitem {maciej}B\l aszak, M. \emph{Theory of separability of
multi-Hamiltonian chains, }J. Math. Phys. \textbf{40 }(1999) 5725
\end{thebibliography}
\end{document}